\documentclass[11pt]{article}
\usepackage{amsfonts}
\usepackage{graphicx}
\usepackage{amsmath}
\usepackage{amssymb}
\usepackage{caption2}
\begin{document}
\bibliographystyle{prsty}
\begin{center}
{\large {\bf \sc{ Analysis of the  scalar  and axial-vector heavy diquark states with QCD sum rules }}} \\[2mm]
Zhi-Gang Wang  \footnote{E-mail:wangzgyiti@yahoo.com.cn. } \\
  Department of Physics, North China Electric Power University, Baoding 071003, P. R.
  China
\end{center}

\begin{abstract}
In this article, we study the mass spectrum of the  scalar and
axial-vector heavy diquark
 states with the QCD sum rules in a
systematic way.   Once the reasonable values are obtained, we can
take them as basic parameters and study the new charmonium-like
states as the tetraquark states.
\end{abstract}


\section{Introduction}
The scattering amplitude for one-gluon exchange in an $SU(N_c)$
gauge theory is proportional to
\begin{eqnarray}
T^a_{ki}
T^a_{lj}&=&-\frac{N_c+1}{4N_c}(\delta_{jk}\delta_{il}-\delta_{ik}\delta_{jl})
 +\frac{N_c-1}{4N_c}(\delta_{jk}\delta_{il}+\delta_{ik}\delta_{jl})\, ,
\end{eqnarray}
where the $T^a$ is the generator of the gauge group, and the $i,j$
and $k,l$ are the color indexes of the two quarks in the incoming
and outgoing channels respectively.   For $N_c=3$, the negative sign
in front of the antisymmetric  antitriplet indicates the interaction
is attractive, while the positive sign in front of the symmetric
sextet indicates
 the interaction  is repulsive \cite{Huang-2005}. The attractive
 interaction favors the formation of the diquark states in the color
 antitriplet, and  the most stable diquark states maybe exist
in  the color antitriplet $\overline{\bf{3}}_{ c}$, flavor
antitriplet $\overline{\bf{3}}_{ f}$ and spin singlet ${\bf{1}}_s$
channels due to Fermi-Dirac statistics \cite{Color-Spin}.  On the
other hand, the study based on the  random instanton liquid model
indicates that the instanton induced quark-quark interactions are
weakly repulsive in the vector and axial-vector channels, strongly
repulsive in the pseudoscalar channel, and strongly attractive in
the scalar and tensor channels \cite{RILM1994}.

 The conception of diquarks has many phenomenological applications
 \cite{diquark-early,ReviewScalar1,CSC-review},
for example,  the quark-diquark bound states scenario of the ground
state baryons \cite{quark-diquark},
 the $\Delta I = \frac{1}{2}$
rule in the non-leptonic weak decays \cite{Delta12}, the  $4:1$
ratio of the proton and neutron deep inelastic structure functions
in the limit $x \to 1$ \cite{Close41}, the color superconductivity
in the cold dense quark matters  \cite{CSC-original}.

 Taking the diquarks as basic constituents, we can obtain a new spectroscopy for the mesons
and baryons \cite{Jaffe1977,DiquarkM}. The numerous candidates  with
$J^{PC}=0^{++}$ below $2\,\rm{GeV}$ cannot be accommodated in one
$q\bar{q}$ nonet, some are supposed to be glueballs, molecular
states and tetraquark states
\cite{ReviewScalar1,Close2002,ReviewScalar2}. The $a(980)$ and
$f(980)$ are good candidates for the $K{\bar K}$ molecular states
\cite{IsgurKK}, however,  their cousins  $\sigma(600)$ and
$\kappa(800)$ lie considerably higher than the corresponding
thresholds, it is difficult to identify them as the $\pi\pi$ and
$\pi K$  molecular states respectively. There maybe different
dynamics dominating the $0^{++}$ mesons  below and above
$1\,\rm{GeV}$ which result in two scalar nonets below $1.7\,\rm{
GeV}$. The strong attractions between the diquark states
$(qq)_{\overline{3}}$ and $(\bar{q}\bar{q})_{3}$ in $S$-wave may
result in a nonet tetraquark states  manifested below $1\,\rm{GeV}$.
The conventional $^3P_0$ $\bar{q}q$ nonet would have masses about
$(1.2-1.6) \,\rm{GeV}$, and the well established $^3P_1$ and $^3P_2$
$\bar{q}q$ nonets with $J^{PC}=1^{++}$ and $2^{++}$ respectively lie
in the same region. Furthermore, there are enough candidates for the
$^3P_0$ $\bar{q}q$ nonet mesons, $a_0(1450)$, $f_0(1370)$,
$K^*(1430)$, $f_0(1500)$ and $f_0(1710)$
\cite{ReviewScalar1,Close2002,ReviewScalar2}.

  In the tetraquark
models, the structures of the nonet scalar mesons in the ideal
mixing limit can be symbolically written  as
\begin{eqnarray}
\sigma(600)=ud\bar{u}\bar{d},\;&&f_0(980)={us\bar{u}\bar{s}+
ds\bar{d}\bar{s}\over\sqrt{2}}, \nonumber\\
a_0^-(980)=ds\bar{u}\bar{s},\;&&a_0^0(980)={us\bar{u}\bar{s}
-ds\bar{d}\bar{s}\over\sqrt{2}},\;a_0^+(980)=us\bar{d}\bar{s},
\nonumber\\
\kappa^+(800)=ud\bar{d}\bar{s},\;&&\kappa^0(800)=ud\bar{u}\bar{s},\;\;\;\;\;\;
\bar{\kappa}^0(800)=us\bar{u}\bar{d},
\;\;\;\;\;\;\kappa^-(800)=ds\bar{u}\bar{d}. \nonumber
\end{eqnarray}
 The four light
isospin-$\frac{1}{2}$ $K\pi$ resonances near $800 \,\rm{MeV}$, known
as the $\kappa(800)$  mesons,  have not been firmly established yet,
there are still controversy   about their existence due to the large
width and nearby $K\pi$ threshold \cite{PDG}. The E791 collaboration
observed a low-mass scalar $K\pi$ resonance with the Breit-Wigner
mass $(797\pm 19 \pm 43) \,\rm{MeV}$ and width $(410\pm 43\pm87)
\,\rm{MeV}$ in the decay $D^+ \to K^- \pi^+ \pi^+$ \cite{E791}, and
the BES collaboration observed a clear low mass enhancement in the
invariant $K\pi$ mass distribution in the decay $J/\psi \to
\bar{K}^*(892)K^+\pi^-$ with the Breit-Wigner mass $(878 \pm
23^{+64}_{-55}) \,\rm{MeV}$ and width $(499 \pm
52^{+55}_{-87})\,\rm{ MeV}$ \cite{BES-2005}. Recently, the BES
collaboration reported the charged $\kappa(800)$ in the decay
 $J/\psi \to K^*(892)^{\mp} K_s \pi^\pm$ with the Breit-Wigner mass
$(826\pm49_{-34}^{+49}) \,\rm{MeV}$ and width
$(449\pm156_{-81}^{+144})\,\rm{MeV}$ \cite{BES-2008}.

In general, we may expect constructing the tetraquark currents and
studying  the   nonet scalar  mesons below $1\,\rm{GeV}$ as the
tetraquark states with the  QCD sum rules approach
\cite{SVZ79,Reinders85}, which is  powerful  tool in studying the
ground state hadrons. For the conventional mesons and baryons, the
"single-pole + continuum states" model works well in representing
the phenomenological spectral densities, the continuum states are
usually approximated by the contributions from the asymptotic quarks
and gluons, the Borel windows  are rather large and reliable QCD sum
rules can be obtained. However, for the light flavor tetraquark
states (and pentaquark states, for example, the $\Theta(1540)$), we
cannot obtain a Borel window to satisfy the two criteria (pole
dominance and convergence of the operator product expansion) of the
QCD sum rules \cite{Narison04}. For the heavy tetraquark states and
molecular states, the two criteria can be satisfied, but the Borel
windows are rather small \cite{WangTetraquark,Matheus2007}.

We can take the colored diquarks as point particles and describe
them as the scalar, pseudoscalar, vector, axial-vector and tensor
fields respectively to overcome the embarrassment \cite{Wang1003}.
In Ref.\cite{Wang1008},  we construct the color singlet tetraquark
currents with the scalar diquark fields, take the diquark masses as
the basic parameters, parameterize the nonperturbative effects with
the new vacuum condensates  besides the gluon condensate,  and
perform the standard procedure of the QCD sum rules to study the
nonet scalar mesons below $1\,\rm{GeV}$, the numerical results are
satisfactory.

In recent years, the Babar, Belle, CLEO, D0, CDF and FOCUS
collaborations have discovered (or confirmed) a large number of
charmonium-like states, and revitalized  the interest in the
spectroscopy of the charmonium states \cite{Recent-review}. For
example, Maiani et al identify the $X(3872)$ observed  by the Belle
collaboration \cite{Belle-3872} as a $1^{++}$ state with the
diquark-antidiquark structure  $[cq]_{S=1}[{\bar c}{\bar
q}]_{S=0}+[cq]_{S=0}[{\bar c}{\bar q}]_{S=1}$ \cite{Maiani-3872}.
 The $Z(4430)$ and $Z(4050)$, $Z(4250)$ observed in the
  decay modes $\psi^\prime\pi^+$ and $\chi_{c1}\pi^+$ respectively by the
 Belle collaboration are the most
interesting subjects \cite{Belle-z4430,Belle-z4430-PRD,Belle-chipi}.
 They can't be  pure $c\bar{c}$ states
due to the positive charge,  and  may be  $c\bar{c}u\bar{d}$
tetraquark states (irrespective of the molecule type and the
diquark-antidiquark type).

In Refs.\cite{diquark-SR,HuangDiquark}, the mass spectrum of the
 scalar light diquark states  are studied using the QCD sum rules. So it
is interesting to study the  scalar and axial-vector heavy diquark
states with the QCD sum rules.  Once reasonable values of the heavy
diquark masses are obtained, we can take them as basic parameters
and study the new charmonium-like states as the tetraquark states.

There have been several theoretical approaches to deal with the
heavy diquark masses, such as the relativistic quark model based on
a quasipotential approach in QCD \cite{Ebert0512}, the
Bethe-Salpeter equation \cite{BS-diquark}, the constituent diquark
model \cite{Maiani-3872,Polosa0902,b-diqaurk}, etc.

The article is arranged as follows:  we derive the QCD sum rules for
the   scalar and axial-vector heavy diquark states  in Sect.2; in
Sect.3, we present the
 numerical results and discussions; and Sect.4 is reserved for our
conclusions.

\section{ The   scalar and axial-vector heavy diquark states  with  QCD Sum Rules}
In the following, we write down the interpolating currents  for the
 scalar and axial-vector heavy  diquark states,
\begin{eqnarray}
J^i(x)&=&\epsilon^{ijk} q_j^T(x)C\gamma_5 Q_k(x)\, , \nonumber\\
\eta^i(x)&=&\epsilon^{ijk} s_j^T(x)C\gamma_5 Q_k(x)\, ,\nonumber\\
J^{i}_\mu(x)&=&\epsilon^{ijk} q_j^T(x)C\gamma_\mu Q_k(x)\, , \nonumber\\
\eta^{i}_\mu(x)&=&\epsilon^{ijk} s_j^T(x)C\gamma_\mu Q_k(x)\, ,
\end{eqnarray}
the $i,~j,~k$ are color indexes, $Q=b,~c$, and the $C$ is the charge
conjugation matrix.

The  two-point correlation functions $\Pi(p)$ and $\Pi_{\mu\nu}(p)$
can be written as
\begin{eqnarray}
 \Pi(p)&=&i\int d^4x ~e^{ip.x}\langle 0|T[J(x)J^\dagger(0)]|0\rangle \, ,  \nonumber\\
 \Pi_{\mu\nu}(p)&=&i\int d^4x ~e^{ip.x}\langle 0|T[J_\mu(x)J_\nu^\dagger(0)]|0\rangle \, ,
\end{eqnarray}
where the  currents  $J(x)$ and $J_\mu(x)$ denote the $J^i(x)$,
$\eta^i(x)$ and $J^i_\mu(x)$, $\eta^i_{\mu}(x)$ respectively.

The one-gluon exchange results in strong attractions in the color
antitriplet  channel $\bar{\bf{3}}_c$, the quark-quark system maybe
form quasibound states (diquark states) which are characterized  by
the correlation length $\mathbb{L}$. At the distance $l>\mathbb{L}$,
the  $\bar{\bf{3}}_c$ diquark state combines with the one quark or
one $\bf{3}_c$ antidiquark to form a baryon state or a tetraquark
state, while at the distance $l<\mathbb{L}$, the
 $\bar{\bf{3}}_c$ diquark states dissociate into asymptotic quarks and gluons gradually.
 We carry out the operator product expansion at not so deep
 Euclidean space where the approximation of the correlation functions by the perturbative contributions  and the
  vacuum condensates makes sense. If we take the diquark state as an effective colored hadron and the diquark mass as an
  effective quantity, $M_\mathbb{D}\sim \frac{1}{\mathbb{L}}$,
    the  correlation function can be continued  to the physical region, where
  the quark-quark correlations exist,
  although the diquarks are not asymptotic states, there are significant differences  between the diquark states and
  conventional hadrons.  The correlation functions are approximated by a pole term plus a
 perturbative continuum.

We can insert  a complete set of intermediate "hadronic" states
(effective hadron states) with the same quantum numbers as the
current operators into the correlation functions    to obtain the
"hadronic" representation \cite{SVZ79}. Isolating the ground state
contributions from the pole terms of the scalar and axial-vector
heavy diquarks, we get the results,
\begin{eqnarray}
\Pi(p)&=&\int_{\Delta^2}^{\infty}ds\frac{\rho_S(s)}{s-p^2} \, , \nonumber\\
\Pi_{\mu\nu}(p)&=&\int_{\Delta^2}^{\infty}ds\frac{\rho_A(s)}{s-p^2}\left(-g_{\mu\nu}+\frac{p_\mu
p_\nu}{p^2}\right)+\cdots \, ,\\
\rho_S(s)&=&\lambda_S^2\delta(s-M_S^2)+\frac{{\rm{Im}}\Pi(s)}{\pi}|_{\rm{per}}\Theta(s-s_0)\, , \nonumber\\
\rho_A(s)&=&\lambda_A^2\delta(s-M_A^2)+\frac{{\rm{Im}}\Pi_{\mu\nu}(s)}{\pi}|_{\rm{per}}\Theta(s-s_0)\,
,
\end{eqnarray}
where the pole residues $\lambda_S$ and $\lambda_A$ (to be more
precise, the current-diquark coupling strengths) are defined as
\begin{eqnarray}
 \langle 0 | {J/\eta}^i(0)|S^j(p)\rangle &=&\lambda_S \delta_{ij}\, , \nonumber\\
 \langle 0 | {J/\eta}^i_\mu(0)|A^j(p)\rangle &=&\lambda_A\epsilon_\mu \delta_{ij} \, ,
 \end{eqnarray}
the $\epsilon_\mu$ is the polarization vector, the $\Delta^2$ is the
threshold parameter,  the $s_0$ is the continuum threshold
parameter, and the $\rm{per}$ denotes the perturbative contributions
in the operator product expansion.

In the following, we briefly outline the operator product expansion
for the correlation functions in perturbative QCD. The calculations
are performed at large space-like momentum region $p^2\ll 0$, which
corresponds to small distance $x\approx 0$ required by   validity of
operator product expansion. We write down the "full" propagators
$S_{ij}(x)$ and $S_Q^{ij}(x)$ of a massive quark in the presence of
the vacuum condensates firstly \cite{Reinders85},
\begin{eqnarray}
S_{ij}(x)&=& \frac{i\delta_{ij}\!\not\!{x}}{ 2\pi^2x^4}
-\frac{\delta_{ij}m_s}{4\pi^2x^2}-\frac{\delta_{ij}}{12}\langle
\bar{s}s\rangle +\frac{i\delta_{ij}}{48}m_s
\langle\bar{s}s\rangle\!\not\!{x}-\frac{\delta_{ij}x^2}{192}\langle \bar{s}g_s\sigma Gs\rangle\nonumber\\
&& +\frac{i\delta_{ij}x^2}{1152}m_s\langle \bar{s}g_s\sigma
 Gs\rangle \!\not\!{x}-\frac{i}{32\pi^2x^2} g_sG^{ij}_{\mu\nu} (\!\not\!{x}
\sigma^{\mu\nu}+\sigma^{\mu\nu} \!\not\!{x})  +\cdots \, ,\nonumber\\
S_Q^{ij}(x)&=&\frac{i}{(2\pi)^4}\int d^4k e^{-ik \cdot x} \left\{
\frac{\delta_{ij}}{\!\not\!{k}-m_Q}
-\frac{g_sG^{\alpha\beta}_{ij}}{4}\frac{\sigma_{\alpha\beta}(\!\not\!{k}+m_Q)+(\!\not\!{k}+m_Q)
\sigma_{\alpha\beta}}{(k^2-m_Q^2)^2}\right.\nonumber\\
&&\left.+\frac{\pi^2}{3} \langle \frac{\alpha_sGG}{\pi}\rangle
\delta_{ij}m_Q \frac{k^2+m_Q\!\not\!{k}}{(k^2-m_Q^2)^4}
+\cdots\right\} \, ,
\end{eqnarray}
where $\langle \bar{s}g_s\sigma Gs\rangle=\langle
\bar{s}g_s\sigma_{\alpha\beta} G^{\alpha\beta}s\rangle$  and
$\langle \frac{\alpha_sGG}{\pi}\rangle=\langle
\frac{\alpha_sG_{\alpha\beta}G^{\alpha\beta}}{\pi}\rangle$, then
contract the quark fields in the correlation functions
 with Wick theorem, and obtain the result:
\begin{eqnarray}
\Pi(p)&=&-i \epsilon^{ijk}\epsilon^{ij'k'} \int d^4x \, e^{i p \cdot
x} Tr\left\{ \gamma_5 S_Q^{jj'}(x)\gamma_5 C
S^T_{kk'}(x)C\right\}\, , \nonumber \\
\Pi_{\mu\nu}(p)&=&i \epsilon^{ijk}\epsilon^{ij'k'} \int d^4x \, e^{i
p \cdot x} Tr\left\{ \gamma_\mu S_Q^{jj'}(x)\gamma_\nu C
S^T_{kk'}(x)C\right\}\, .
\end{eqnarray}
Substitute the full $s$, $c$ and $b$ quark propagators into above
correlation functions and complete  the integral in  coordinate
space, then integrate over the variable $k$, we can obtain the
correlation functions  at the level of quark-gluon degrees  of
freedom.
  Once  the analytical  expressions are obtained,
  then we can take the dualities below the thresholds
$s_0$ (the  continuum states above the thresholds $s_0$ are
asymptotic quarks and the spectral densities on both sides coincide)
and perform the Borel transform with respect to the variable
$P^2=-p^2$, finally we obtain the following sum rules for the heavy
diquark states contain one $s$ quark,
\begin{eqnarray}
\lambda_k^{2}e^{-\frac{M_k^{2}}{M^2}}&=&\int_{\Delta_k^2}^{s^k_0}ds
e^{-\frac{s}{M^2}}\rho_k(s)\, ,
\end{eqnarray}
where the subscript (or superscript) $k$ denotes the scalar and
axial-vector channels, i.e. $k=S$, $A$,
\begin{eqnarray}
\rho_{S}(s)&=&\frac{3}{2\pi^2}\int_\alpha^1 dx
\left[x(1-x)(3s-2\widetilde{m}_Q^2) +m_sm_Q\right]-2m_Q\langle
\bar{s}s\rangle\delta(s-m_Q^2)\nonumber\\
&&+m_s\langle\bar{s}s\rangle\left( 1+\frac{m_Q^2}{M^2}\right)\delta(s-m_Q^2)+\frac{m_Q^3\langle\bar{s}g_s\sigma Gs\rangle}{2M^4}\delta(s-m_Q^2)\nonumber\\
&&-\frac{m_sm_Q^4\langle\bar{s}g_s\sigma
Gs\rangle}{6M^6}\delta(s-m_Q^2)-\frac{m_Q^4}{12M^4}\langle\frac{\alpha_sGG}{\pi}\rangle\int_0^1dx\frac{1-x}{x^3}
\delta(s-\widetilde{m}_Q^2)  \nonumber \\
&&+\frac{m_sm_Q}{6M^2}\langle\frac{\alpha_sGG}{\pi}\rangle\int_0^1dx\frac{1}{x^2}\left(1-\frac{m_Q^2}{2M^2}\right)
\delta(s-\widetilde{m}_Q^2)  \nonumber \\
&&+\frac{1}{8}\langle\frac{\alpha_sGG}{\pi}\rangle\int_0^1dx\left(1+\frac{s}{M^2}\right)
\delta(s-\widetilde{m}_Q^2) \, ,
\end{eqnarray}
\begin{eqnarray}
\rho_{A}(s)&=&\frac{3}{2\pi^2}\int_\alpha^1 dx
\left[x(1-x)(2s-\widetilde{m}_Q^2) +m_sm_Q\right]-2m_Q\langle
\bar{s}s\rangle\delta(s-m_Q^2)\nonumber\\
&&+\frac{m_sm_Q^2\langle\bar{s}s\rangle }{M^2}\delta(s-m_Q^2)+\frac{m_Q^3\langle\bar{s}g_s\sigma Gs\rangle}{2M^4}\delta(s-m_Q^2)\nonumber\\
&&+\frac{m_s\langle\bar{s}g_s\sigma
Gs\rangle}{6M^2}\left(1+\frac{m_Q^2}{M^2}-\frac{m_Q^4}{M^4}\right)\delta(s-m_Q^2)\nonumber\\
 &&+\frac{m_Q^2}{12M^2}\langle\frac{\alpha_sGG}{\pi}\rangle\int_0^1dx\frac{1-x}{x^2}\left(1-\frac{s}{M^2}
 \right)\delta(s-\widetilde{m}_Q^2)  \nonumber \\
&&+\frac{m_sm_Q}{6M^2}\langle\frac{\alpha_sGG}{\pi}\rangle\int_0^1dx\frac{1}{x^2}\left(1-\frac{m_Q^2}{2M^2}\right)
\delta(s-\widetilde{m}_Q^2)  \nonumber \\
&&-\frac{1}{12}\langle\frac{\alpha_sGG}{\pi}\rangle\int_0^1dx\left(1+\frac{s}{2M^2}\right)
\delta(s-\widetilde{m}_Q^2) \, ,
\end{eqnarray}
 $\widetilde{m}_{Q}^2=\frac{m_{Q}^2}{x}$, $\alpha=\frac{m_Q^2}{s}$,
  $\Delta_k^2=(m_Q+m_s)^2$,   and the $M^2$ is the Borel
  parameter. In Eqs.(9-11), we use the dispersion relation to write the spectral densities in a compact form,
   the integrals of the types $\int_{\Delta_k^2}^{s^k_0}ds f(s)e^{-\frac{s}{M^2}}\delta(s-m_Q^2)$ and
    $\int_{\Delta_k^2}^{s^k_0}ds f(s)e^{-\frac{s}{M^2}}\delta(s-\widetilde{m}_Q^2)$ should be carried out formally, i.e.
    $\int_{\Delta_k^2}^{s^k_0}ds f(s)e^{-\frac{s}{M^2}}\delta(s-m_Q^2)=f(m_Q^2)e^{-\frac{m_Q^2}{M^2}}$ and
    $\int_{\Delta_k^2}^{s^k_0}ds f(s)e^{-\frac{s}{M^2}}\delta(s-\widetilde{m}_Q^2)=f(\widetilde{m}_Q^2)e^{-\frac{\widetilde{m}_Q^2}{M^2}}$ despite
    the values $\Delta_k^2>m_Q^2$ and $\Delta_k^2\geq$ or $< \, \widetilde{m}_Q^2$, where the $f(s)$ denotes the spectral densities concerning the vacuum condensates.   With a simple replacement $m_s\rightarrow0$,
  $\langle\bar{s}s\rangle\rightarrow\langle\bar{q}q\rangle$, $\langle\bar{s}g_s\sigma Gs\rangle\rightarrow
  \langle\bar{q}g_s\sigma Gq\rangle$, we can
  obtain the corresponding sum rules for the heavy diquark states contain one
  $q$ quark.

   Differentiate  Eq.(9) with respect to  $\frac{1}{M^2}$, then eliminate the
 pole residues $\lambda_{k}$, we can obtain the sum rules for
 the diquark masses,
 \begin{eqnarray}
 M_{k}^2=\frac{ \int_{\Delta_{k}^2}^{s_0^{k}} ds \frac{d}{d(-1/M^2)}
\rho_k(s)e^{-\frac{s}{M^2}} }{\int_{\Delta_{k}^2}^{s_0^{k}} ds
\rho_k(s)e^{-\frac{s}{M^2}}}\, .
\end{eqnarray}

\section{Numerical Results}

The input parameters are taken to be the standard values $\langle
\bar{q}q \rangle=-(0.24\pm 0.01 \,\rm{GeV})^3$, $\langle \bar{s}s
\rangle=(0.8\pm 0.2 )\langle \bar{q}q \rangle$, $\langle
\bar{q}g_s\sigma Gq \rangle=m_0^2\langle \bar{q}q \rangle$, $\langle
\bar{s}g_s\sigma Gs \rangle=m_0^2\langle \bar{s}s \rangle$,
$m_0^2=(0.8 \pm 0.2)\,\rm{GeV}^2$,  $m_s=(0.14\pm0.01)\,\rm{GeV}$,
$m_c=(1.35\pm0.10)\,\rm{GeV}$ and $m_b=(4.7\pm0.1)\,\rm{GeV}$ at the
energy scale  $\mu=1\, \rm{GeV}$ \cite{SVZ79,Reinders85,Ioffe2005}.

The $Q$-quark masses appearing in the perturbative terms  are
usually taken to be the pole masses in the QCD sum rules, while the
choice of the $m_Q$ in the leading-order coefficients of the
higher-dimensional terms is arbitrary \cite{NarisonBook,Kho9801}.
The $\overline{MS}$ mass $m_c(m_c^2)$ relates with the pole mass
$\hat{m}_c$ through the relation $ m_c(m_c^2)
=\hat{m}_c\left[1+\frac{C_F \alpha_s(m_c^2)}{\pi}+\cdots\right]^{-1}
$. In this article, we take the approximation
$m_c(m_c^2)\approx\hat{m}_c$ without the $\alpha_s$ corrections for
consistency. The value listed in the  Review of Particle Physics is
$m_c(m_c^2)=1.27^{+0.07}_{-0.11} \, \rm{GeV}$ \cite{PDG}, it is
reasonable to take
$\hat{m}_c=m_c(1\,\rm{GeV}^2)=(1.35\pm0.10)\,\rm{GeV}$. For the $b$
quark,  the $\overline{MS}$ mass is
$m_b(m_b^2)=4.20^{+0.17}_{-0.07}\,\rm{GeV}$ \cite{PDG}, the
  gap between the energy scale $\mu=4.2\,\rm{GeV}$ and
 $1\,\rm{GeV}$ is rather large, the approximation $\hat{m}_b\approx m_b(m_b^2)\approx m_b(1\,\rm{GeV}^2)$ seems rather crude.
  It would be better to understand the quark masses $m_c$ and $m_b$ we
take at the energy scale $\mu^2=1\,\rm{GeV}^2$ as the effective
quark masses (or just the mass parameters). Our previous works on
the mass spectrum of the heavy and doubly heavy baryon states
indicate such parameters can lead to satisfactory results
\cite{Wang-Baryon}.

In the conventional QCD sum rules \cite{SVZ79,Reinders85}, there are
two criteria (pole dominance and convergence of the operator product
expansion) for choosing  the Borel parameter $M^2$ and threshold
parameter $s_0$. In practice, we usually  consult the experimental
data in choosing those parameters.

Here we take a short digression to illustrate the two  criteria of
the QCD sum rules. The pole contributions (or the ratios of the pole
contributions) $R_k$ are defined by
\begin{eqnarray}
R_k&=&\frac{\int_{\Delta_k^2}^{s^k_0}ds
e^{-\frac{s}{M^2}}\rho_k(s)}{\int_{\Delta_k^2}^{\infty}ds
e^{-\frac{s}{M^2}}\rho_k(s)} \, ,
\end{eqnarray}
for a definite channel $k$ at the hadronic representation, and the
pole dominance requires $R_k\geq 50\%$, i.e. the pole contributions
dominate over the continuum contributions.  At the level of
quark-gluon degrees of freedom,  the convergence of the operator
product expansion requires the operators of increasing dimension of
mass (for example, $1$, $3$, $4$, $5$, $\cdots$) should have smaller
contributions.

 If we multiply the Bethe-Salpeter amplitudes of the diquark
states by a charge conjunction matrix $C$, the scalar and
axial-vector diquark states have the same Bethe-Salpeter equation as
the pseudoscalar and vector mesons respectively, except for the
interacting kernels have an additional factor $\frac{1}{2}$
\cite{BS-SA}, the  scalar and axial-vector diquark states maybe have
slightly larger (or equal) masses than (or of) that of the
corresponding pseudoscalar and vector mesons respectively. In the
QCD sum rules for the conventional mesons and baryons, we usually
 take the energy gap between the ground
states and the first radial excited states to be $0.5\,\rm{GeV}$. We
 take the approximation $M_S\approx M_P$ and $M_A\approx M_V$,
and  determine the central values of the threshold parameters
tentatively, $s^0_S=(M_P+0.5)^2\,\rm{GeV}$ and
$s^0_A=(M_V+0.5)^2\,\rm{GeV}$, where the $S$ and $A$ denote the
scalar and axial-vector diquark states respectively, the $P$ and $V$
denote the corresponding pseudoscalar  and vector mesons
respectively.

In calculation, we take analogous pole contributions and uniform
Borel windows, i.e. $M^2_{max}-M^2_{min}=1.0\,\rm{GeV}^2$ and
$1.5\,\rm{GeV}^2$ in the charmed and bottom channels respectively,
where the platforms are rather flat. The revelent parameters are
shown explicitly in Table 1, from the Table, we can see that the
pole contributions are in the range $(45-86)\%$, the contributions
from the different terms in the operator product expansion have the
hierarchy: perturbative-term $\gg\langle\bar{q}q\rangle+
\langle\bar{q}g_s\sigma Gq\rangle\gg \langle
\frac{\alpha_sGG}{\pi}\rangle$, the two criteria (pole dominance and
convergence of the operator product expansion) are well satisfied.
Taking into account all uncertainties of the relevant parameters, we
can obtain the values of the masses and pole resides of  the scalar
and axial-vector heavy diquark states, which are shown in Tables
2-3.   From  Table 2, we can see that the scalar and axial-vector
diquark states have almost degenerate masses with the corresponding
pseudoscalar and vector mesons respectively. We should bear in mind
that those values are not necessarily the lowest masses.

In this article, we intend to obtain the possible lowest masses,
which correspond to the largest correlation lengths,  and impose the
two criteria of the QCD sum rules on the scalar and axial-vector
heavy diquark states to choose the Borel parameter $M^2$ and
threshold parameter $s_0$, i.e. we take smaller threshold parameters
and Borel parameters (also Borel windows)  than that presented in
Table 1, and adjust them to warrant the uniform pole contributions
(about $(46-71)\%$, the smallest pole contribution presented in
Table 1). The preferred values are shown in Fig.1 and Table 4, where
we can see explicitly that the pole contributions are about
$(45-70)\%$, and the contributions from the different terms in the
operator product expansion have the hierarchy: perturbative-term
$>\langle\bar{q}q\rangle+ \langle\bar{q}g_s\sigma Gq\rangle\gg
\langle \frac{\alpha_sGG}{\pi}\rangle$, the two criteria of the QCD
sum rules are well satisfied also. On the other hand, the values of
the masses and pole residues are rather stable with variations of
the Borel parameters in the Borel windows. In fact, we can take even
 smaller threshold parameters than that presented in Table 4, however, the
  Borel windows are too small to make reliable predictions.

Taking into account all uncertainties of the relevant  parameters,
finally we obtain the (lowest) values of the masses and pole resides
of the scalar and axial-vector heavy diquark states, which are shown
in Figs.2-3 and Tables 2-3. In Table 2, we also present the masses
from the relativistic quark model based on a quasipotential approach
in QCD \cite{Ebert0512}, the Bethe-Salpeter equation
\cite{BS-diquark}, and the constituent diquark model
\cite{Maiani-3872,Polosa0902,b-diqaurk}. From the table,  we can see
that the values from different theoretical approaches differ  from
each other greatly,  and one should be careful when using them. In
Ref.\cite{Wang1008}, we introduce new QCD sum rules to study the
nonet scalar mesons  and take the values of the scalar diquark
masses from the QCD sum rules for consistency \cite{HuangDiquark}.

The $SU(3)$ breaking effects for the masses of the scalar and
axial-vector heavy diquark states are buried in the uncertainties.
Naively, we expect the axial-vector heavy diquark states have larger
masses than the corresponding scalar heavy diquark states. From
Table 2, we can see that it is not the case, they have degenerate
masses.  Lattice QCD calculations for the light flavors  indicate
that the strong attraction in the scalar diquark channels favors the
formation of  good diquarks, the weaker attraction  in the
axial-vector diquark channels maybe form bad diquarks, the energy
gap between the axial-vector and scalar diquarks is about
$\frac{2}{3}$ of the $\Delta$-nucleon mass splitting, i.e. $\approx
0.2\,\rm{GeV}$ \cite{Latt-1,Latt-2}, which is expected from the
hypersplitting color-spin  interaction
$\frac{C}{m_im_j}\vec{T}_{i}\cdot \vec{T}_{j} \vec{\sigma}_i \cdot
\vec{\sigma}_j$, where the $C$ is a coefficient
\cite{Color-Spin,ReviewScalar1}. The coupled rainbow Dyson-Schwinger
equation and ladder Bethe-Salpeter equation also indicate such an
energy hierarchy \cite{BS-diquark-light}. Comparing with the light
diquark states, the contribution from the hypersplitting color-spin
interaction $\frac{C}{m_im_j}\vec{T}_{i}\cdot \vec{T}_{j}
\vec{\sigma}_i \cdot \vec{\sigma}_j$  to the heavy diquark states is
greatly suppressed due to the large constituent quark masses, and
the scalar and axial-vector heavy diquark states have almost
degenerate masses.

\begin{table}
\begin{center}
\begin{tabular}{|c|c|c|c|c|c|c|c|}
\hline\hline & $M^2 (\rm{GeV}^2)$& $s_0 (\rm{GeV}^2)$&pole&perturbative & $\langle \bar{q}q\rangle+\langle \bar{q}g_s\sigma Gq\rangle$ & $\langle \frac{\alpha_sGG}{\pi}\rangle$\\
\hline
 $cq(0^+)$  &$1.2-2.2$ &$5.6\pm0.3$&  $(45-86)\%$&$(59-73)\%$ & $(26-40)\%$ &$<1\%$\\ \hline
 $cq(1^+)$  &$1.7-2.7$ &$6.3\pm0.3$&  $(45-78)\%$&$(68-78)\%$ &$(23-33)\%$ &$<1\%$\\ \hline
  $cs(0^+)$  &$1.4-2.4$ &$6.1\pm0.3$&  $(46-83)\%$&$(76-85)\%$ & $(15-23)\%$ &$<1\%$\\ \hline
 $cs(1^+)$  &$2.0-3.0$ &$6.8\pm0.3$&  $(46-74)\%$&$(82-87)\%$ & $(13-19)\%$ &$<1\%$\\ \hline
  $bq(0^+)$  &$3.8-5.3$ &$33.0\pm1.0$&  $(46-76)\%$&$(64-71)\%$ & $(29-36)\%$ &$<1\%$\\ \hline
  $bq(1^+)$  &$4.5-6.0$ &$34.0\pm1.0$&  $(46-72)\%$&$(68-74)\%$ & $(26-32)\%$ &$<1\%$\\ \hline
  $bs(0^+)$  &$4.5-6.0$ &$34.5\pm1.0$&  $(46-72)\%$&$(80-84)\%$ & $(16-20)\%$ &$<1\%$\\ \hline
  $bs(1^+)$  &$4.9-6.4$ &$35.0\pm1.0$&  $(46-71)\%$&$(81-85)\%$ & $(16-20)\%$ &$<1\%$\\ \hline
  \hline
\end{tabular}
\end{center}
\caption{ The  Borel parameters $M^2$ and threshold parameters $s_0$
for the  heavy diquark states. The "pole" stands for the
contribution from the pole term to the spectral density. The
"perturbative" stands for the contribution from the perturbative
term in the operator product expansion, etc, where the central value
of the threshold parameter $s_0$ is taken. The contributions from
the $\langle\bar{q}q\rangle$ and $\langle\bar{q}g_s\sigma Gq\rangle$
are not shown independently  for simplicity, and
 $\langle\bar{q}q\rangle\gg \langle\bar{q}g_s\sigma Gq\rangle$.}
\end{table}

\begin{table}
\begin{center}
\begin{tabular}{|c|c|c|c|c|c|c|}
\hline\hline
    &$\widehat{M} $&  $M$ &  Ref.\cite{Ebert0512}&Refs.\cite{Maiani-3872,Polosa0902,b-diqaurk}& Ref.\cite{BS-diquark}&Ref.\cite{PDG}\\ \hline
   $cq(0^+)$ &$1.86\pm0.10$&$1.77\pm0.08$ & 1.793 & 1.933& 2.088 &1.867 $[D]$\\ \hline
   $cq(1^+)$ &$1.96\pm0.10$& $1.76\pm0.08$ & 2.036 & & 2.067 & 2.009 $[D^*]$\\ \hline
   $cs(0^+)$  &$1.98\pm0.10$&$1.84\pm0.08$ & 2.091 & 1.955& 2.192&1.969 $[D_s]$\\ \hline
   $cs(1^+)$  &$2.08\pm0.09$&$1.84\pm0.08$ & 2.158 & &2.168&2.112 $[D^*_s]$\\ \hline
   $bq(0^+)$  &$5.23\pm0.09$&$5.14\pm0.12$ & 5.359 & 5.267 &5.556 &5.279 $[B]$\\ \hline
   $bq(1^+)$  &$5.28\pm0.09$&$5.13\pm0.11$ & 5.381 & & 5.539&5.325 $[B^*]$\\ \hline
   $bs(0^+)$  &$5.35\pm0.09$&$5.20\pm0.07$ & 5.462 & &5.648&5.366 $[B_s]$\\ \hline
   $bs(1^+)$  &$5.38\pm0.09$&$5.20\pm0.08$ & 5.482 & &5.636&5.415 $[B^*_s]$\\ \hline
  \hline
\end{tabular}
\end{center}
\caption{ The masses $M$ of the scalar and axial-vector heavy
diquark states, the values are in unit of $\rm{GeV}$, the wide-hat
denotes the values from the parameters presented in Table 1.    }
\end{table}

\begin{table}
\begin{center}
\begin{tabular}{|c|c|c|}
\hline\hline
    &$\widehat{\lambda} $&  $\lambda $ \\ \hline
   $cq(0^+)$ & $0.53\pm0.08$  &$0.43\pm0.05$\\ \hline
   $cq(1^+)$ & $0.57\pm0.08$   &$0.40\pm0.04$\\ \hline
   $cs(0^+)$  & $0.64\pm0.09$ &$0.48\pm0.05$\\ \hline
   $cs(1^+)$  & $0.69\pm0.08$ &$0.45\pm0.05$\\ \hline
   $bq(0^+)$  & $1.02\pm0.14$ &$0.78\pm0.14$\\ \hline
   $bq(1^+)$  &$1.08\pm0.14$ &$0.76\pm0.12$\\ \hline
   $bs(0^+)$  &$1.27\pm0.17$ &$0.91\pm0.14$\\ \hline
   $bs(1^+)$  &$1.28\pm0.17$ &$0.88\pm0.14$\\ \hline
  \hline
\end{tabular}
\end{center}
\caption{ The  pole residues $\lambda$ of the scalar and
axial-vector heavy diquark states,  the values are in unit of
$\rm{GeV}^2$, the wide-hat denotes the values from the parameters
presented in Table 1.  }
\end{table}

\begin{table}
\begin{center}
\begin{tabular}{|c|c|c|c|c|c|c|c|}
\hline\hline & $M^2 (\rm{GeV}^2)$& $s_0 (\rm{GeV}^2)$&pole&perturbative & $\langle \bar{q}q\rangle+\langle \bar{q}g_s\sigma Gq\rangle$ & $\langle \frac{\alpha_sGG}{\pi}\rangle$\\
\hline
 $cq(0^+)$  &$1.2-1.8$ &$4.8\pm0.2$&  $(46-76)\%$&$(54-62)\%$ & $(37-45)\%$ &$<1\%$\\ \hline
 $cq(1^+)$  &$1.3-1.9$ &$4.8\pm0.2$&  $(47-75)\%$&$(53-60)\%$ &$(42-50)\%$ &$<3\%$\\ \hline
  $cs(0^+)$  &$1.2-1.9$ &$5.0\pm0.2$&  $(45-77)\%$&$(69-76)\%$ & $(23-30)\%$ &$<1\%$\\ \hline
 $cs(1^+)$  &$1.3-2.0$ &$5.0\pm0.2$&  $(46-76)\%$&$(68-74)\%$ & $(27-34)\%$ &$<2\%$\\ \hline
  $bq(0^+)$  &$3.5-4.2$ &$30.0\pm1.0$&  $(45-67)\%$&$(53-55)\%$ & $(45-47)\%$ &$<1\%$\\ \hline
  $bq(1^+)$  &$3.6-4.3$ &$30.0\pm1.0$&  $(46-67)\%$&$(52-54)\%$ & $(46-49)\%$ &$<1\%$\\ \hline
  $bs(0^+)$  &$3.5-4.4$ &$31.0\pm1.0$&  $(46-71)\%$&$(71-73)\%$ & $(27-29)\%$ &$<1\%$\\ \hline
  $bs(1^+)$  &$3.6-4.6$ &$31.0\pm1.0$&  $(45-71)\%$&$(70-72)\%$ & $(28-31)\%$ &$<1\%$\\ \hline
  \hline
\end{tabular}
\end{center}
\caption{ The preferred Borel parameters $M^2$ and threshold
parameters $s_0$ for the  heavy diquark states. The "pole" stands
for the contribution from the pole term to the spectral density. The
"perturbative" stands for the contribution from the perturbative
term in the operator product expansion, etc, where the central value
of the threshold parameter $s_0$ is taken. The contributions from
the $\langle\bar{q}q\rangle$ and $\langle\bar{q}g_s\sigma Gq\rangle$
are not shown independently  for simplicity, and
 $\langle\bar{q}q\rangle\gg \langle\bar{q}g_s\sigma Gq\rangle$. }
\end{table}

\begin{figure}
 \centering
 \includegraphics[totalheight=5cm,width=6cm]{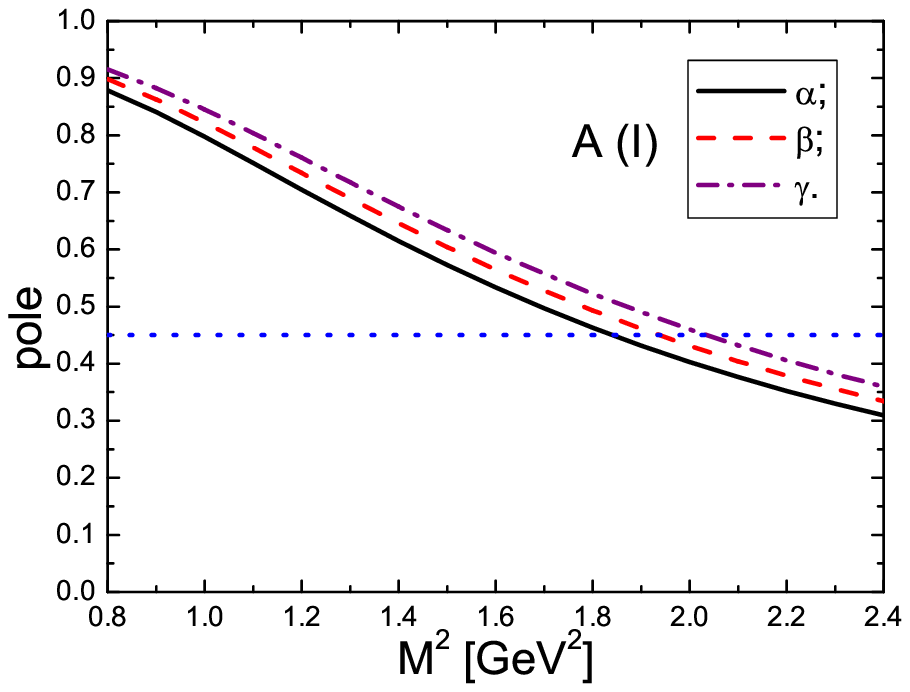}
 \includegraphics[totalheight=5cm,width=6cm]{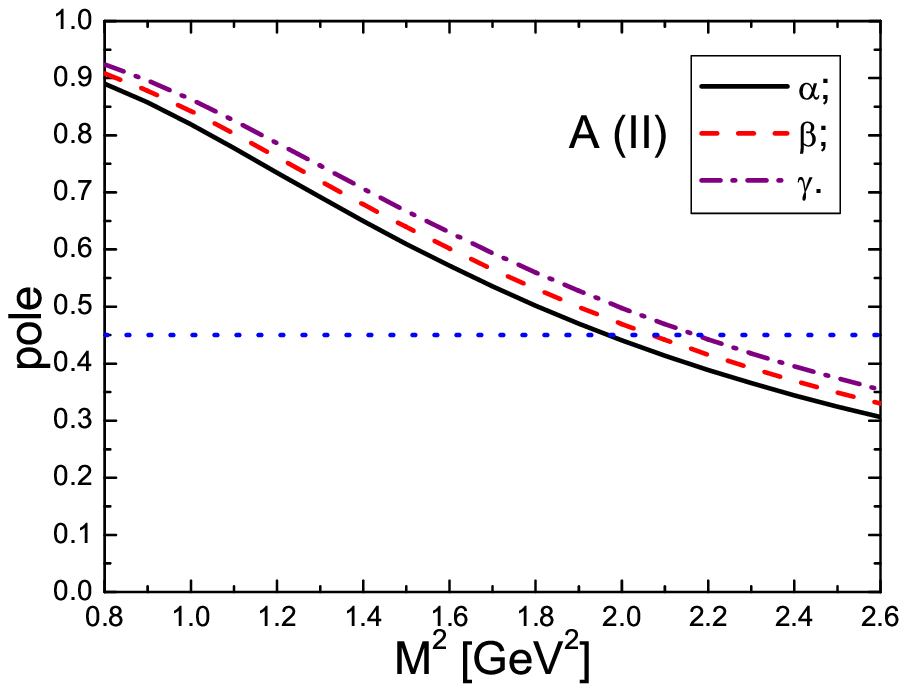}
 \includegraphics[totalheight=5cm,width=6cm]{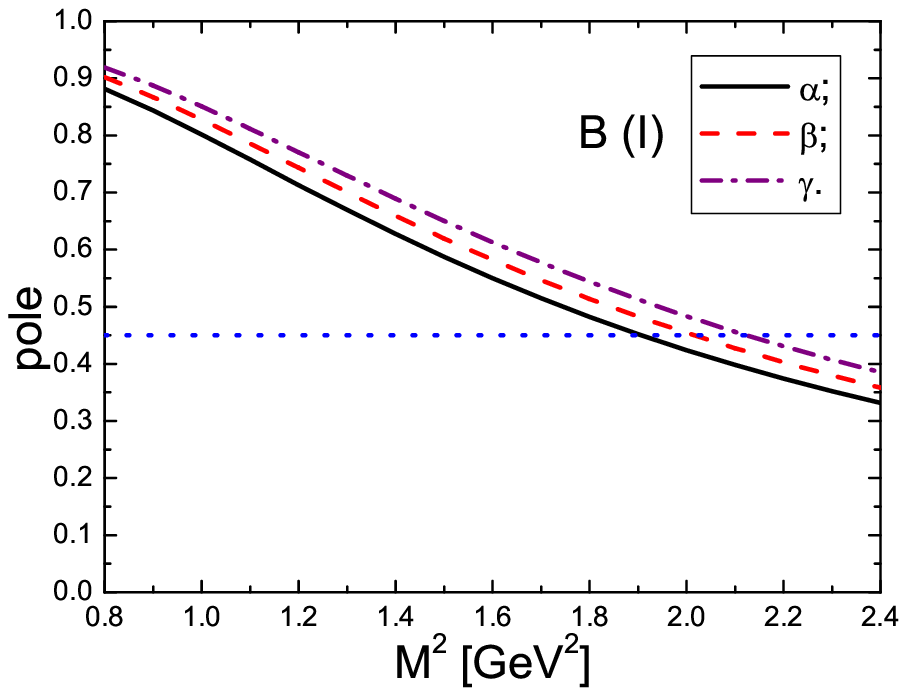}
 \includegraphics[totalheight=5cm,width=6cm]{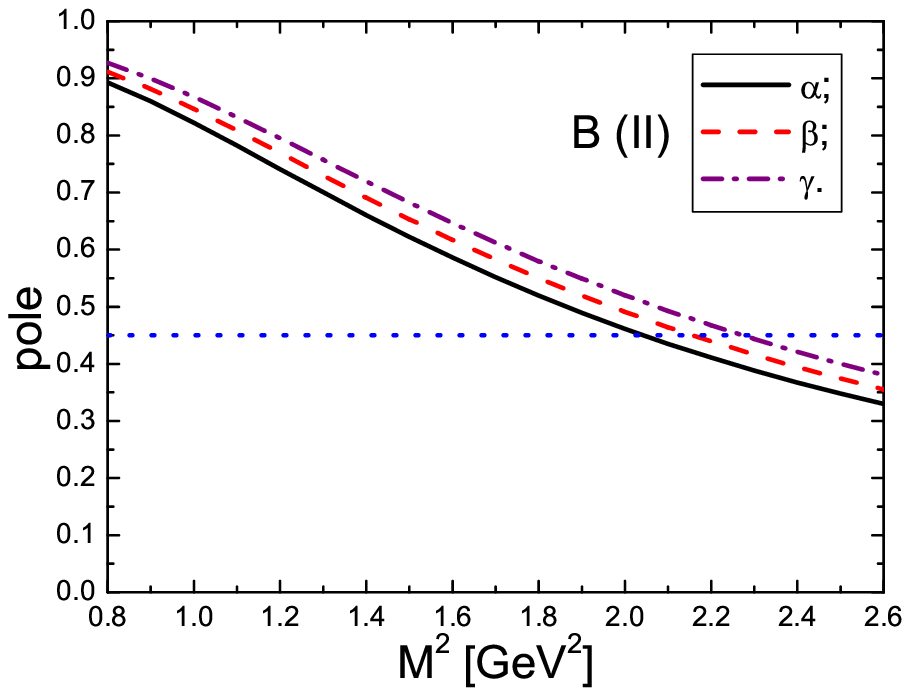}
 \includegraphics[totalheight=5cm,width=6cm]{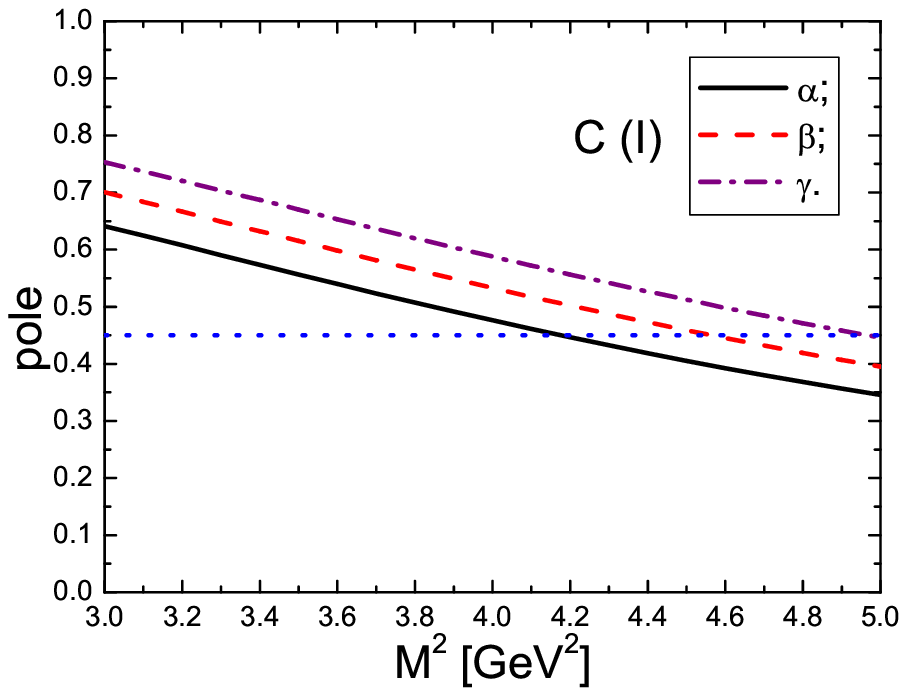}
 \includegraphics[totalheight=5cm,width=6cm]{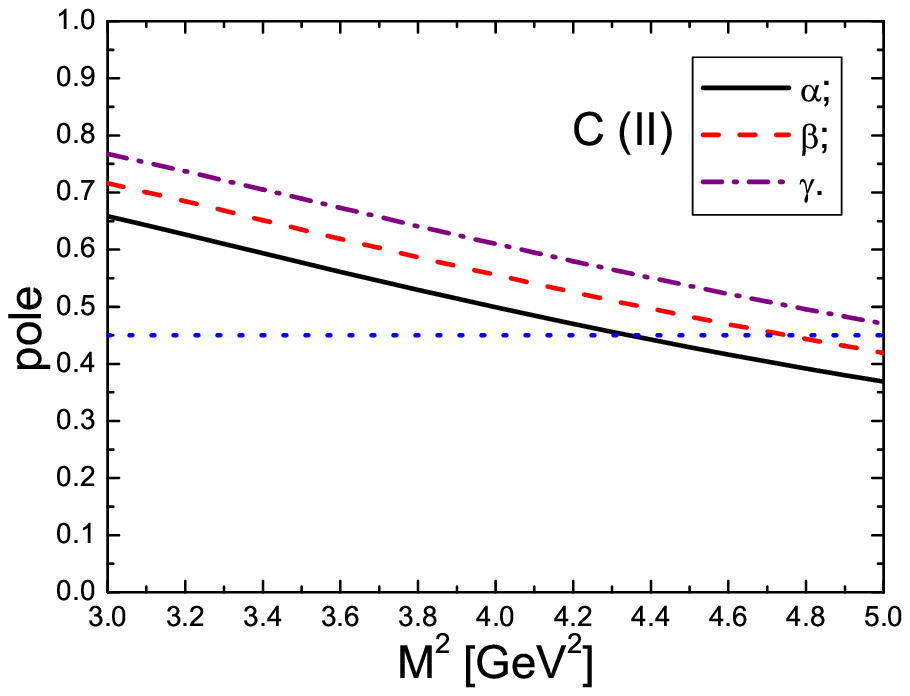}
 \includegraphics[totalheight=5cm,width=6cm]{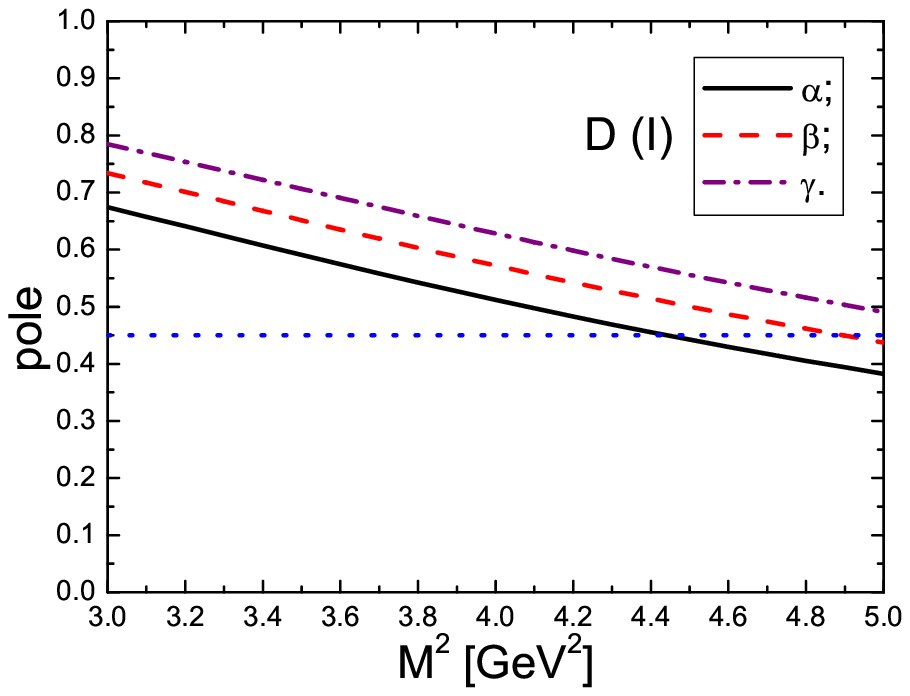}
 \includegraphics[totalheight=5cm,width=6cm]{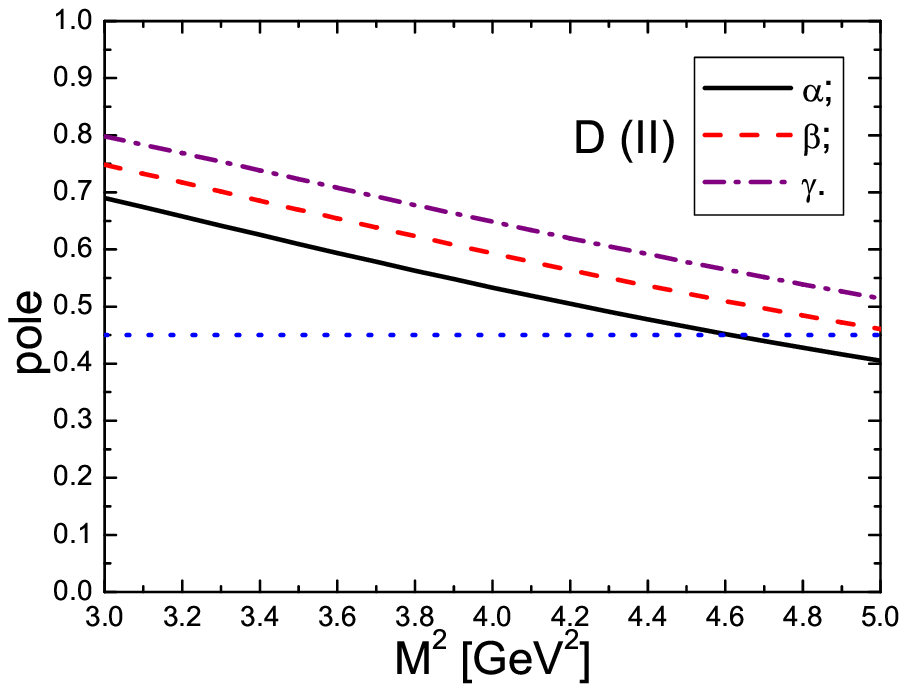}
   \caption{ The  contributions of the pole terms to the spectral densities with variation of the Borel parameter
   $M^2$. The $A$, $B$, $C$ and  $D$ denote the channels $cq$,
   $cs$,  $bq$    and $bs$ respectively.  The $\beta$ corresponds to the
central values of the threshold parameters, the energy gaps among
$\alpha$, $\beta$ and $\gamma$ are $0.2\,\rm{GeV}$ and
$1.0\,\rm{GeV}$ for the charmed and bottom diquark states
respectively. The (I) and (II) denote the scalar and axial-vector
diquark states respectively. }
\end{figure}

\begin{figure}
 \centering
 \includegraphics[totalheight=5cm,width=6cm]{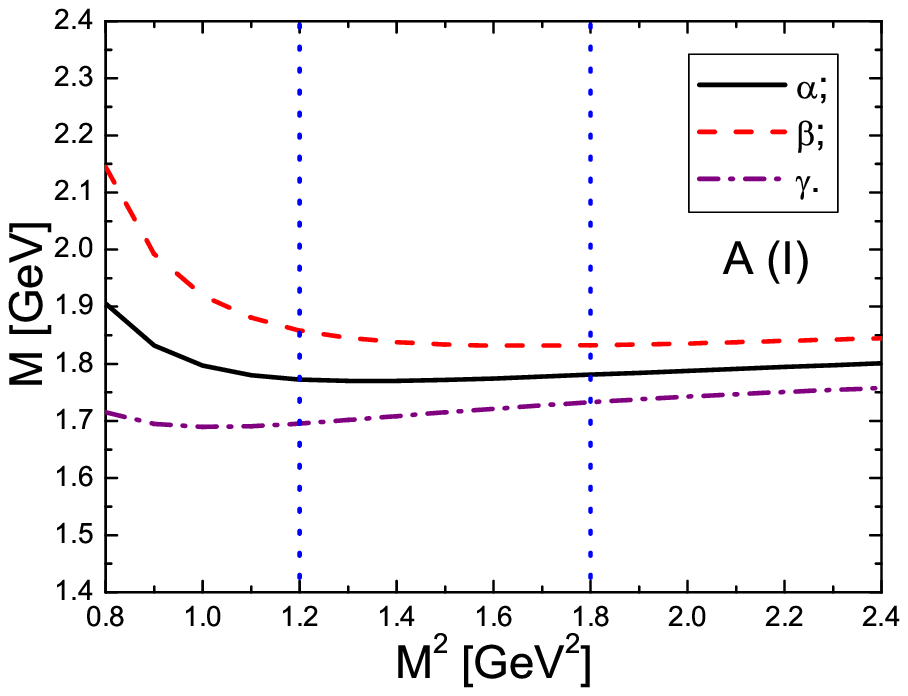}
 \includegraphics[totalheight=5cm,width=6cm]{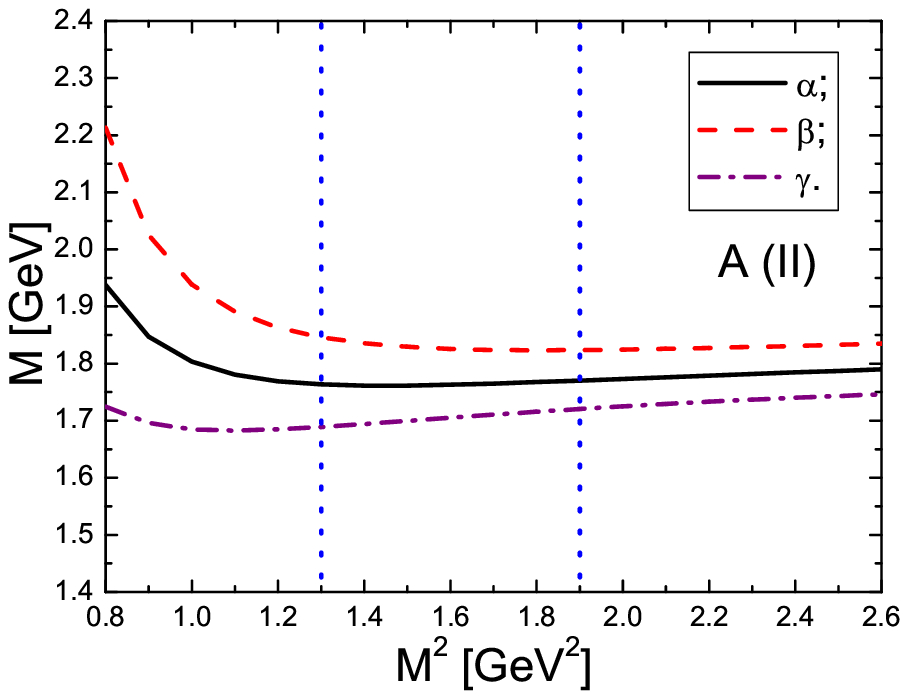}
 \includegraphics[totalheight=5cm,width=6cm]{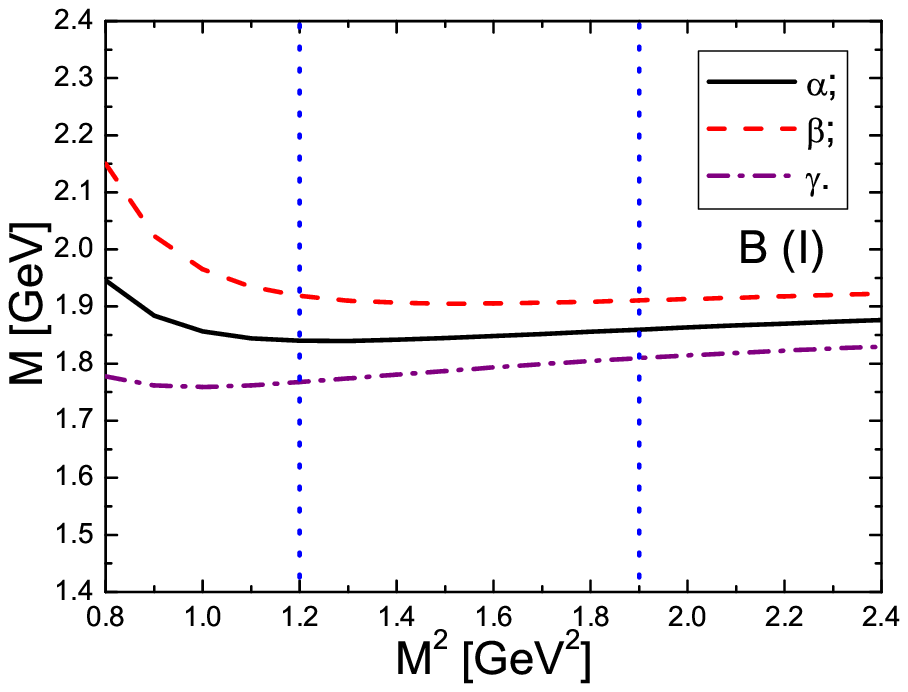}
 \includegraphics[totalheight=5cm,width=6cm]{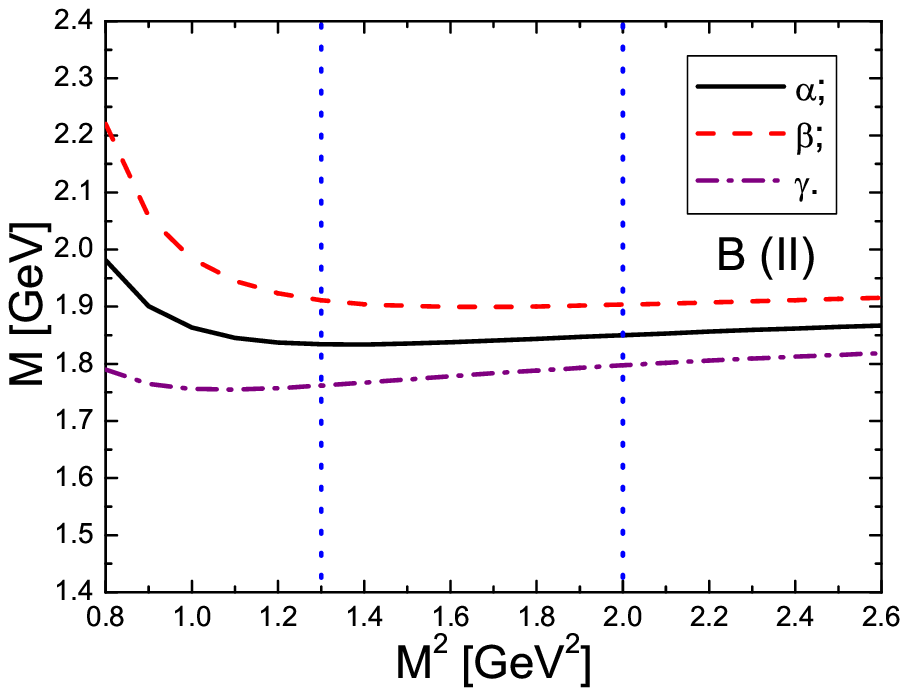}
 \includegraphics[totalheight=5cm,width=6cm]{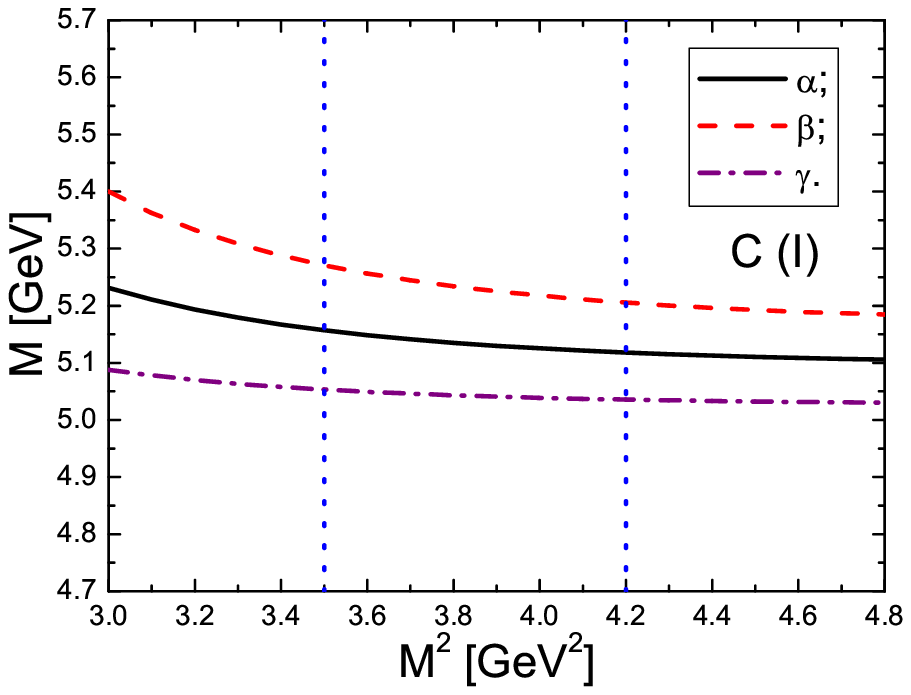}
 \includegraphics[totalheight=5cm,width=6cm]{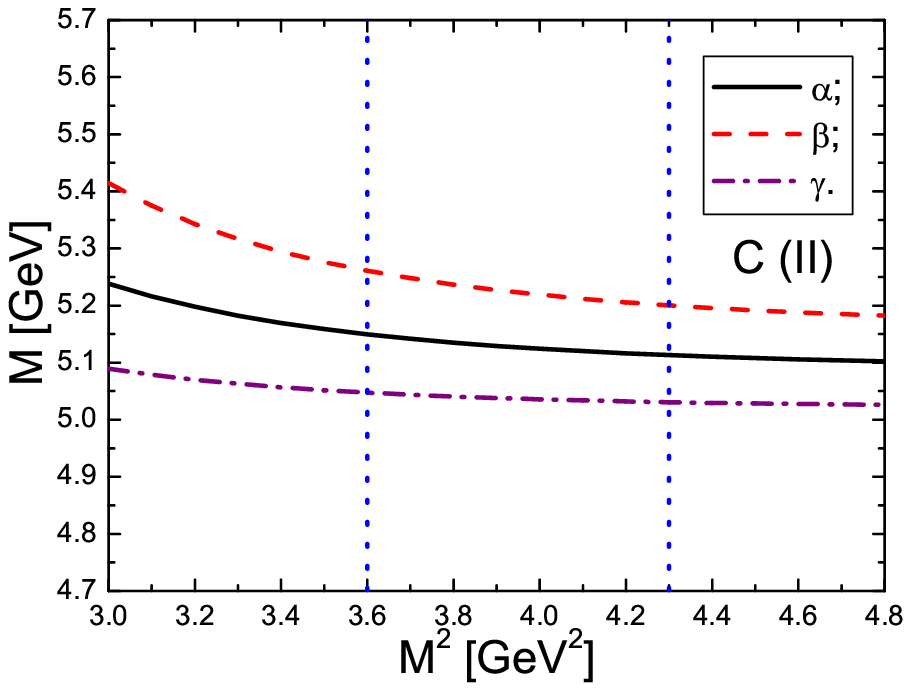}
 \includegraphics[totalheight=5cm,width=6cm]{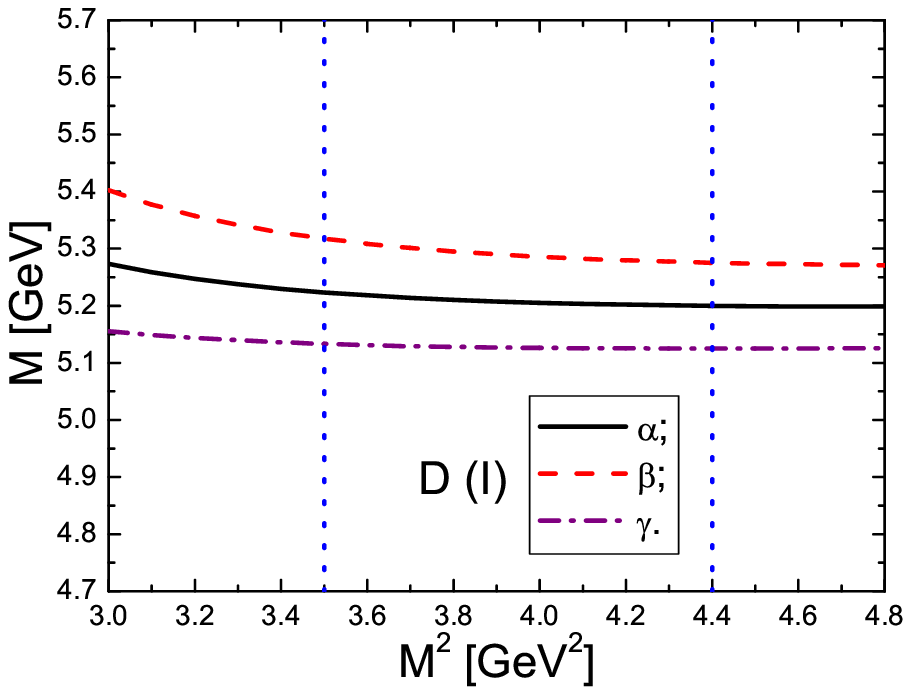}
 \includegraphics[totalheight=5cm,width=6cm]{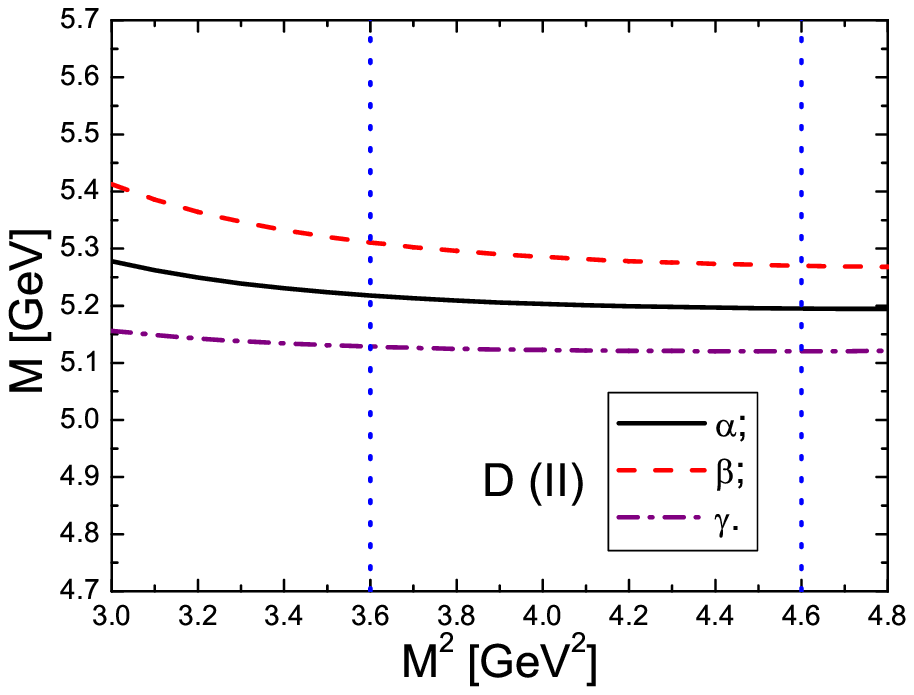}
   \caption{The values of the  diquark masses with variation of the Borel parameter
   $M^2$. The $A$, $B$, $C$ and  $D$ denote the channels $cq$,
   $cs$,  $bq$    and $bs$ respectively.  The $\alpha$, $\beta$ and $\gamma$ denote the upper bound,  the
central value  and the lower bound respectively. The (I) and (II)
denote the scalar and axial-vector diquark states respectively.}
\end{figure}

\begin{figure}
 \centering
 \includegraphics[totalheight=5cm,width=6cm]{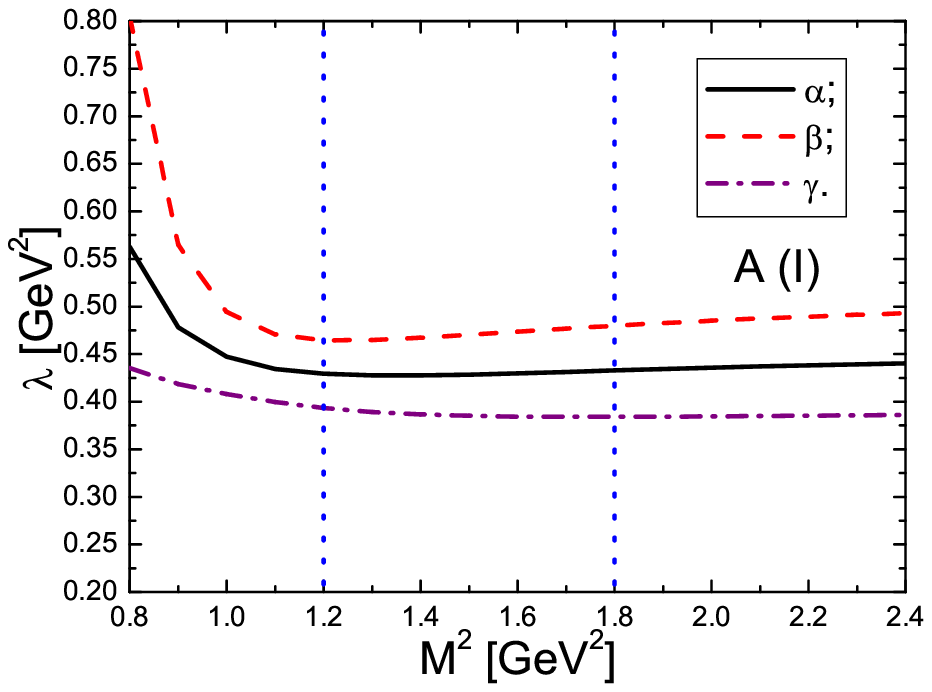}
 \includegraphics[totalheight=5cm,width=6cm]{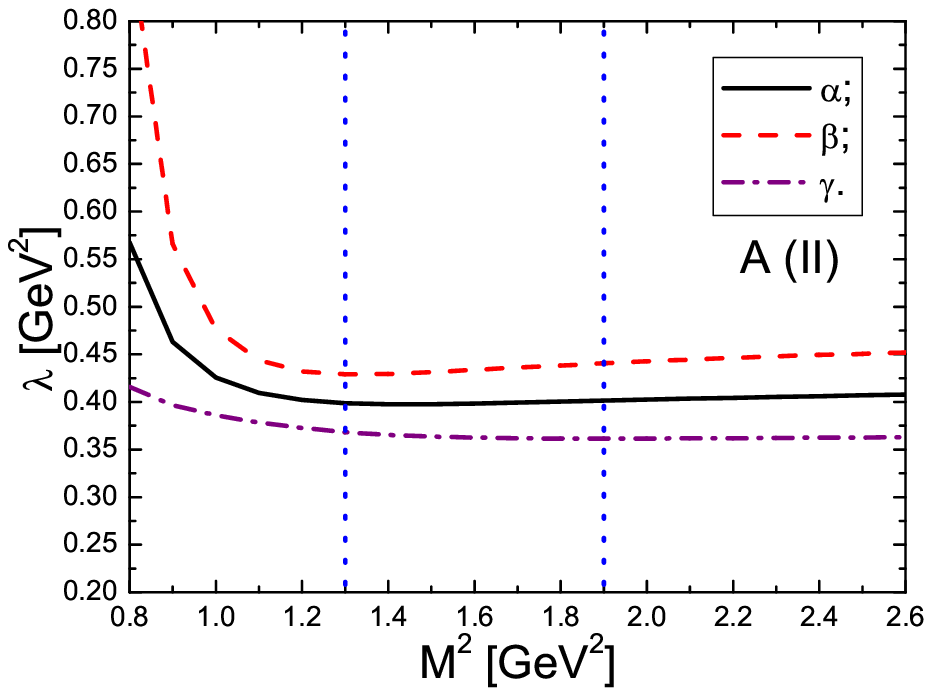}
 \includegraphics[totalheight=5cm,width=6cm]{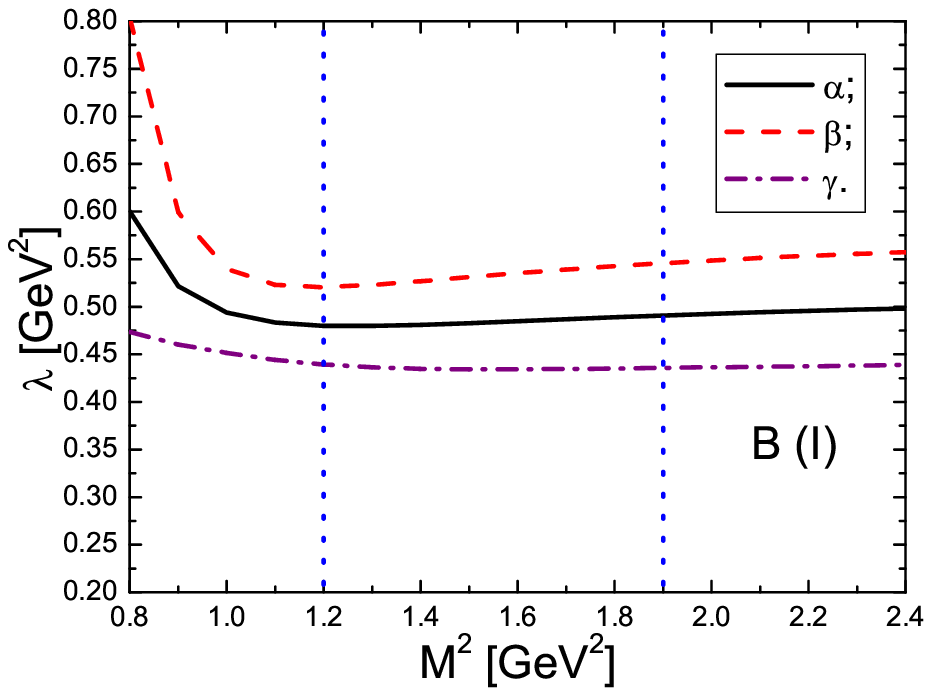}
  \includegraphics[totalheight=5cm,width=6cm]{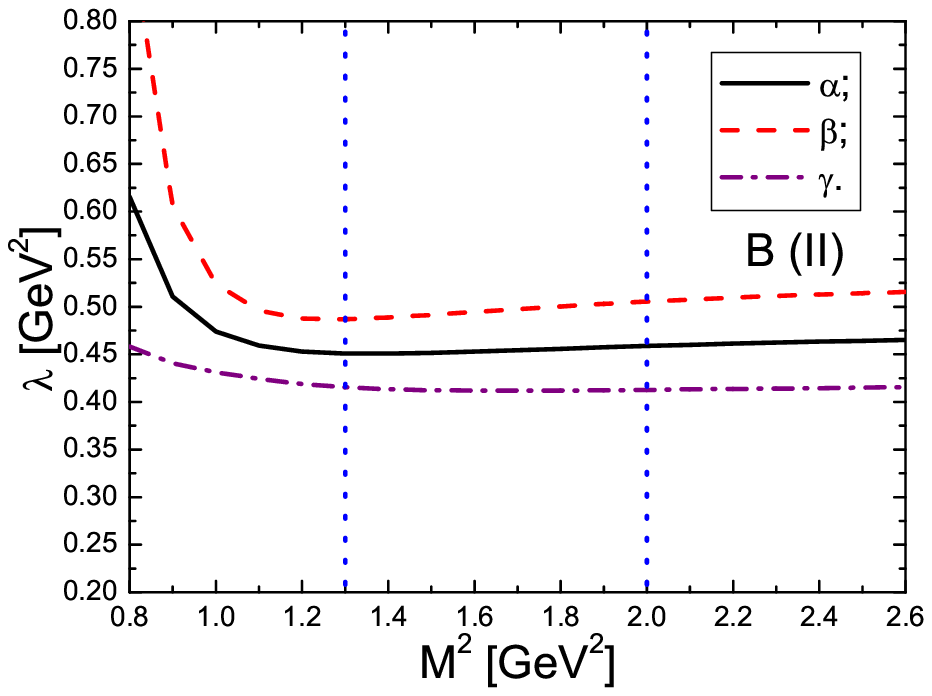}
  \includegraphics[totalheight=5cm,width=6cm]{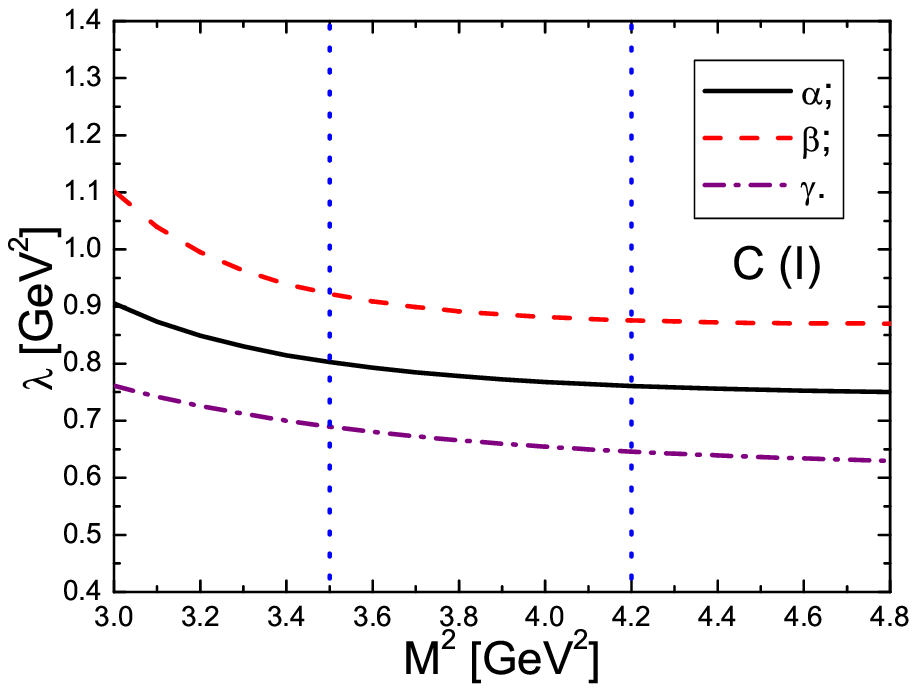}
 \includegraphics[totalheight=5cm,width=6cm]{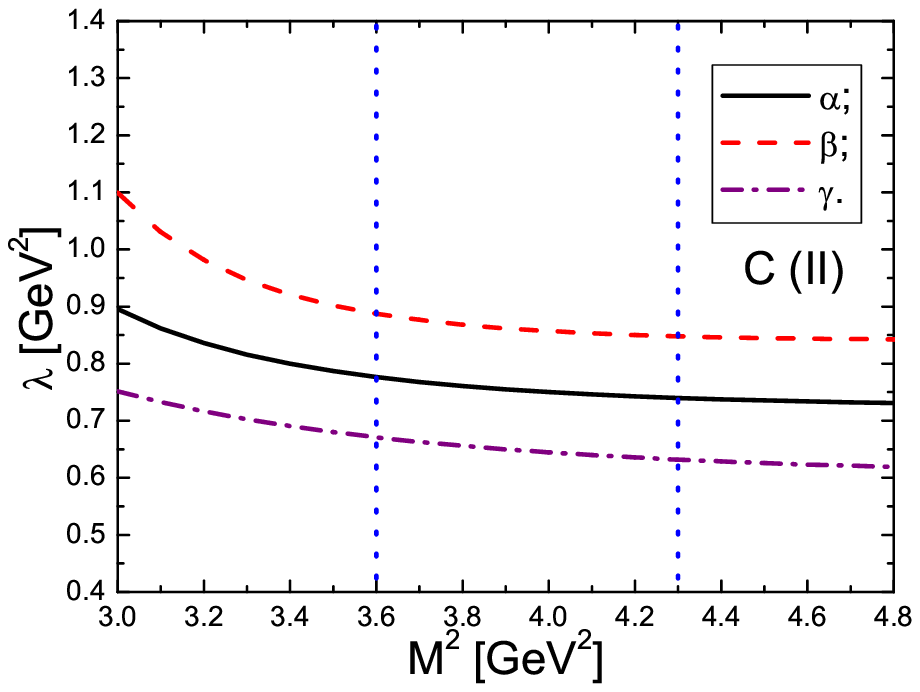}
 \includegraphics[totalheight=5cm,width=6cm]{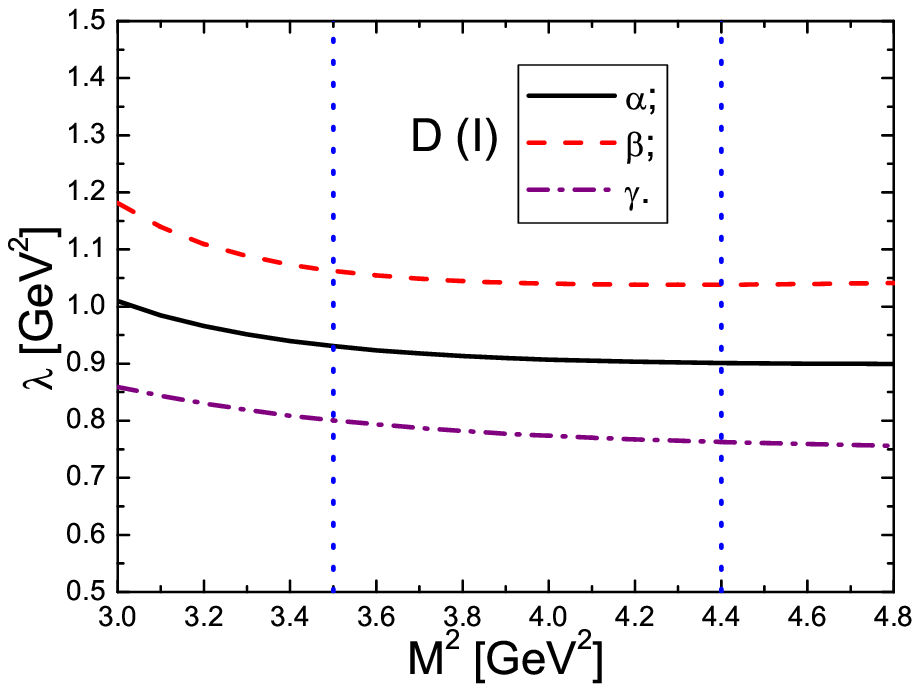}
  \includegraphics[totalheight=5cm,width=6cm]{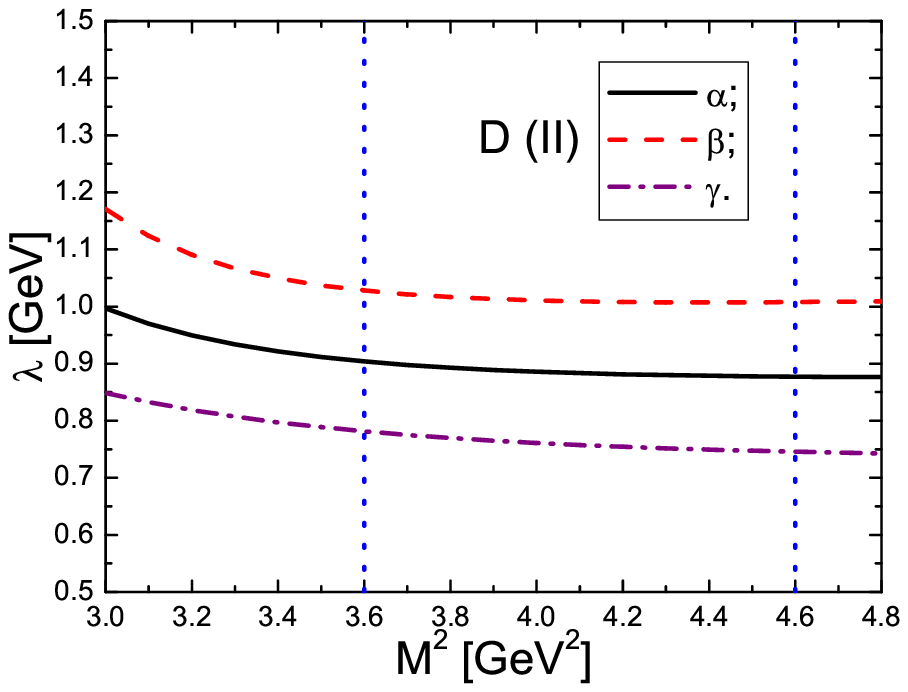}
   \caption{ The values of the  pole residues with variation of the Borel parameter
   $M^2$. The $A$, $B$, $C$ and  $D$ denote the channels $cq$,
   $cs$,  $bq$    and $bs$ respectively.  The $\alpha$, $\beta$ and $\gamma$ denote the upper bound,  the
central value  and the lower bound respectively. The (I) and (II)
denote the scalar and axial-vector diquark states respectively.}
\end{figure}

\section{Conclusion}
In this article, we study the mass spectrum of the  scalar and
axial-vector heavy diquark
 states with the QCD sum rules in a
systematic way.  The diquark masses are basic parameters in studying
the tetraquark states, once reasonable values are obtained, we can
 study the new charmonium-like states as the
tetraquark states with the new QCD sum rules developed in our
previous work.

\section*{Acknowledgment}
This  work is supported by National Natural Science Foundation of
China, Grant Numbers 10775051, 11075053, and Program for New Century
Excellent Talents in University, Grant Number NCET-07-0282, and the
Fundamental Research Funds for the Central Universities.

\end{document}